\documentclass[debug]{rmaa}

\usepackage{psfrag,color}

\usepackage[latin1]{inputenc}

\def\kms{km s$^{-1}$}

\def\hii{H{\sc ii}}
\def\msun{M$_\odot$}

\def\mjyb{mJy beam$^{-1}$}
\def\jyb{Jy beam$^{-1}$}

\def\cmdos{cm$^{-2}$}
\def\cmtres{cm$^{-3}$}
\def\x{$\times$}
\def\gra{$^{\circ}$}

\def\radec{RA,Dec.(J2000)}
\def\dco{$^{12}$CO(2-1)}
\def\tco{$^{13}$CO(2-1)}
\def\cdo{C$^{18}$O(2-1)}
\def\lafuente{G341.3553-00.2885}
\def\etal{et al.~}

\title{A submillimeter study of the IR dust bubble S\,21 and its environs} 

\author{
  C. E. Cappa,\altaffilmark{1,2} 
  N. U. Duronea,\altaffilmark{1}
  J. Vasquez,\altaffilmark{1,2}
  M. Rubio,\altaffilmark{3}
  V. Firpo,\altaffilmark{4,5}
  C.-H. L\'opez-Caraballo,\altaffilmark{6}
  and J. Borissova,\altaffilmark{7,8}}

\altaffiltext{1}{Instituto Argentino de Radioastronom\'ia, CONICET, Argentina.}
\altaffiltext{2}{Facultad de Ciencias Astron\'omicas y Geof\'isicas, Universidad Nacional de la Plata, Paseo del Bosque s/n, 1900, La Plata, Argentina.}
\altaffiltext{3}{Departamento de Astronom\'ia, Universidad de Chile, Chile.}
\altaffiltext{4}{Gemini-CONICYT, Departamento de F\'isica y Astronom\'ia, Universidad de La Serena, La Serena, Chile.}
\altaffiltext{5}{ Gemini Observatory, Southern Operations Center, La Serena, Chile. }
\altaffiltext{6}{Universidad Cat\'olica de Chile, Santiago, Chile. }

\altaffiltext{7}{Instituto de F\'isica y Astronom\'ia, Universidad de Valpara\'iso,  Chile.}
\altaffiltext{8}{ Millennium Institute of Astrophysics (MAS), Santiago, Chile.}

\shortauthor{Cappa et al.}
\shorttitle{Multiwavelength study of S\,21}

\fulladdresses{
\item Nicol\'as U. Duronea, Javier Vasquez and Cristina E. Cappa: Instituto Argentino de Radioastronom\'ia, CONICET, CCT La PLata, C.C.5, 1894, Villa Elisa, Argentina (cristina.elisabet.cappa@gmail.com).
\item Javier Vasquez and Cristina E. Cappa: Facultad de Ciencias Astron\'omicas y Geof\'isicas, Universidad Nacional de La Plata, Argentina.
\item M\'onica Rubio: Departamento de Astronom\'ia, Universidad de Chile, Chile
  (mrubio@das.uchile.cl).
\item Jura Borissova: Instituto de F\'isica y Astronom\'ia, Universidad de Valpara\'iso, Av. Gran Breta\~na 1111, Playa Ancha, Casilla 5030, Chile.
\item Ver\'onica Firpo: Departamento de F\'isica y Astronom\'ia, Universidad de La Serena, La Serena, Chile.
\item C.-H. L\'opex-Caraballo: Universidad Cat\'olica de Chile, Santiago, Chile.
}

\listofauthors{C. E. Cappa, N.U. Duronea, J. Vasquez, M. Rubio, V. firpo, C.-H. L\'opez Caraballo, \& J. Borissova}
\indexauthor{Cappa, E. E.}
\indexauthor{Duronea, N. U.}
\indexauthor{Vasquez, J.}
\indexauthor{Rubio, M.}
\indexauthor{Firpo, V.}
\indexauthor{L\'opez-Caraballo, C.-H.}
\indexauthor{Borissova, J.}

\abstract{Based on the molecular emission in the   $^{12}$CO(2-1) and  $^{13}$CO(2-1) lines,  and the continuum emission in the MIR and FIR towards the S\,21 IR dust bubble, we analyze the physical characteristics of the gas and dust linked to the nebula and the presence of young stellar objects (YSOs) in its environs. 
The line emission reveals a clumpy molecular shell, 1.4 pc in radius, encircling S\,21. The total molecular mass in the shell amounts to 2900 \msun\ and the original ambient density, 2.1\x 10$^3$ \cmtres, indicating that the bubble is evolving in a high density interstellar medium. The image at 24 $\mu$m shows warm dust inside the bubble, while the emission in the range  250 to 870 $\mu$m  reveal cold dust in its outskirts, coincident with the molecular gas. The detection of  radio continuun emission indicates that the bubble is a compact \hii\ region. A search for YSOs using photometric criteria allowed to identify many candidates projected onto the molecular clumps. We analize if the {\it collect and collapse} process has triggered a new generation of stars. }

\resumen{Basados en la emisi\'on molecular en las líneas   $^{12}$CO(2-1) y $^{13}$CO(2-1), y en la emisi\'on  en el continuo en el mediano y lejano infrarrojo hacia la burbuja S\,21, analizamos las características físicas del gas y polvo asociado  con S\,21 y la presencia de objetos estelares j\'ovenes (YSOs) en su entorno. La emisi\'on molecular revela una cáscara   grumosa de 1.4 pc de radio rodeando a S\,21. Su masa molecular es de  2900 \msun\ y la densidad ambiental original en la región, 2.1\x 10$^3$ \cmtres, indicando que la burbuja evoluciona en un medio de alta densidad. La imagen  a 24 $\mu$m muestra polvo tibio dentro de la burbuja, mientras que la emisi\'on en el rango  250 a 870 $\mu$m revelan que hay polvo  frío en la vecindad, coincidente con el gas molecular. La detección de emisión en el continuo de radio indica que S\,21 es una regi\'on \hii\ compacta. Una b\'usqueda de YSOs 
utilizando criterios fotométricos permitió identificar muchos candidatos coincidentes con los grumos moleculares. Se analiza si el proceso de {\it  collect and collapse} ha dado origen a una nueva generaci\'on de estrellas. }

\addkeyword{H~II regions}
\addkeyword{ISM: molecules}
\addkeyword{ISM: individual (S\,21)}
\addkeyword{Stars: protostars}

\begin{document}
\maketitle

\section{Introduction}

Massive (O and B-type) stars have an enormous impact on their surroundings due to their ultraviolet (UV) ionizing radiation and energetic winds. Ionized gas in \hii\ regions produces strong infrared (IR), optical, and thermal radio continuum emission. In addition, ionized gas mixed with heated dust make an \hii\ region bright in thermal IR emission.

A common spatial shape reported for  individual Galactic \hii\ regions at IR wavelengths  is  the ring morphology, or \lq\lq bubble\rq\rq\ seen in projection. A visual examination of the images  at 8.0 $\mu$m from the Galactic Legacy Infrared Survey Extraordinaire (GLIMPSE; Benjamin \etal 2003) allowed the identification of about 600 full or partial IR dust bubbles (IRDBs) in the inner Galactic plane  (Churchwell \etal 2006, 2007) between longitudes from --60\gra\ to +60\gra. Presently, more than 5000 IRDBs have been identified (Simpson et al. 2012). The main characteristics of many of these bubbles were investigated  by several authors (see for example Deharveng \etal 2010; Alexander \etal 2013).  These bubbles  are about 1\arcmin\ - 3\arcmin\ in size, show filamentary appearence, and many of them lie close to  massive stars and coincide (or enclose) classical and ultracompact \hii\ regions. 

At 8 $\mu$m, most of the emission originates in strong features of polycyclic aromatic hydrocarbons (PAH) molecules,  which are considered to be good tracers of warm UV-irradiated photodissociation regions (PDR; Hollenbach \& Tielens 1997).  Since these complex molecules are destroyed in the ionized gas (Povich \etal 2007; Lebouteiller \etal 2007), they delineate the ionization front and indicate  the presence of substantial amounts of molecular gas  surrounding the bubbles. Therefore, these bubbles provide a good insight of  the sculpting influences of the UV photons of massive stars on the molecular clouds where they are born.

The geometry of the IR bubbles is also important for understanding triggered star formation scenarios. Classical models, like the {\it collect and collapse} mechanism (C\&C; Elmegreen \& Lada 1977) and the {\it radiative driven implosion} process (RDI; Lefloch \& Lazareff 1994), suggest that the formation of stars can be triggered by the action of \hii\ regions over their parental molecular environment.  Molecular condensations lying at the border of  many galactic  bubble-shaped  \hii\ regions are then among the most likely sites for stellar births and to look for early stages of star formation (e.g. Deharveng \etal 2008; Zavagno \etal 2010; Brand \etal 2011; Samal \etal 2014). Detailed studies of IR bubbles have shown the presence of young stellar objects (YSOs) in their environments, although the triggered formation scenario not always can be proven (e.g. Dewangan \& Ojha 2013; Alexander \etal 2013).

As part of a project aimed to study and characterize galactic bubbles, we have selected S\,21 from the sample of bubbles reported by  Churchwell \etal (2006) to perform a  study of its molecular and dust environment, and search for  candidates to YSOs in their vecinity. S\,21  is located 10\arcmin\ east from S\,24 (Cappa et al. 2016). 

\begin{figure}
\centering
\includegraphics[width=8cm]{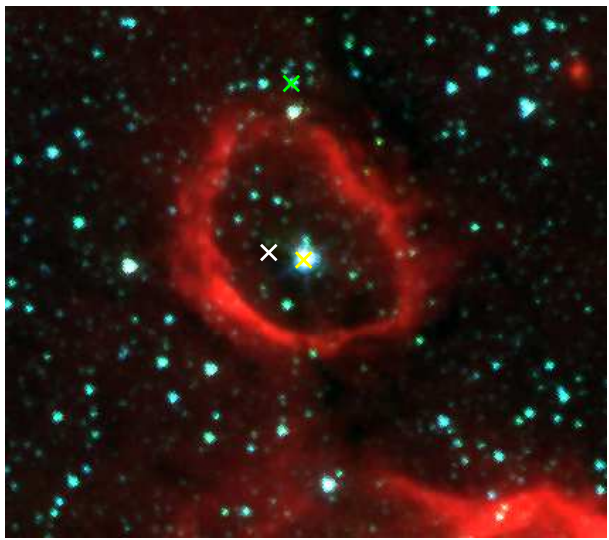}
\caption{Composite IRAC image of S\,21. The emission at 3.6 $\mu$m is in blue, at 4.5 $\mu$m is in green, and at 8.0 $\mu$m is in red. The positions of IRAS\,16495-4418, \lafuente, and the catalogued B3-star are indicated with white, yellow, and green crosses, respectively. }
\label{fig:s21-irac}
\end{figure}

S\,21 is an almost complete circular bubble (see Fig. 1) of $\sim$ 0\farcm 75 in radius centered at \radec\ = 16$^h$53$^m$7.9$^s$, $-$44\gra23\arcmin13\farcs 1. The dusty nebula appears filamentary at 8 $\mu$m, with a rather steep inner border and a more diffuse outer one (Fig.~1). The point source IRAS\,16495-4418 (\radec\ = 16$^h$53$^m$8.5$^s$, $-$44\gra23\arcmin21\arcsec) coincides with the bubble. Watson \etal (2010) determined the dust temperature inside the bubble using the emission at 24 and 70 $\mu$m from Spitzer-MIPS and a modified blackbody. They found a temperature gradient with the highest temperatures (85 K) close to the center of the bubble and  the lowest ones (71 K) close to the border.  

Figure~1 reveals a bright point source detected in the IRAC bands, named G341.3553-00.2885 (\radec\ = 16$^h$53$^m$7.12$^s$, $-$44\gra23\arcmin24\farcs 1), located close to the center of the bubble. It has counterparts in several optical (GSC, NOMAD, and DENIS) and IR catalogues (2MASS, Spitzer, WISE). Projected close to the northern border of the bubble, the B3-type star HD\,329056 (Simbad, 16$^h$53$^m$7.63$^s$, $-$44\gra22\arcmin9\farcs 28) has been identified.

The distance to S\,21 is matter of some debate. Adopting a mean velocity of $-$44 \kms\ for the gas linked to S\,21 (see Sect. 3.1), circular galactic rotation models predict near and far kinematical distances of 3.7 and 12-13 kpc (e.g. Brand \& Blitz 1993). Bearing in mind that most of the IRDBs are closer than 8 kpc (Churchwell \etal 2006), we will adopt  for  S\,21 the near kinematical distance of  3.7 kpc. A distance uncertainty of 0.5 kpc results after taking into account a velocity dispersion of 6 \kms\ for the interstellar gas.

In this study, we present a molecular line and dust continuum analysis  toward  S\,21 and its environs, with the aim of studying the distribution and physical properties (mass, densities, temperature, kinematics, etc.) of the molecular gas and dust associated with the  bubble. We based our study on \dco\ and  \tco\ data obtained with the APEX telescope\footnote{APEX, the Atacama Pathfinder EXperiment, is a collaboration between Max Planck Institut fur Radioastronomie (MPIfR), Onsala Space Observatory (OSO), and the European Southern Observatory (ESO). }, and complementary archival IR, optical,  and radiocontinuum data.

The  simple morphology of S21, along with the strong evidence of star formation  in its environs  (see below),  make this object an excellent laboratory for the investigation of  possible  scenarios of triggered star formation. With that aim, we also analyze the spatial distribution of the candidates to YSOs in their vicinity and their relation to the bubble, and search for probable exciting stars.

\section{Data}

\subsection{Molecular line observations}

The characteristics  of the molecular gas were investigated using \dco\ (at 230.538000 GHz, HPBW = 30\arcsec) and \tco\ (at 220.398677 GHz, HPBW = 28\arcsec) line observations obtained in October 2010 with  the Atacama Pathfinder Experiment (APEX) 12-m telescope (G\''usten \etal 2006) at Llano de Chajnantor, Chile (Project C-086.F-0674B-2010, P.I. M. Rubio). As front end for the observations, we used the APEX-1 receiver of the  Swedish Heterodyne Facility Instrument (SHeFI; Vassilev \etal 2008). The back end for the observations was the eXtended bandwidth Fast Fourier Transform Spectrometer2 (XFFTS2)  with a 2.5 GHz bandwidth divided into 4096 channels.  Under good weather conditions, this leads to APEX-1 DSB system temperatures of about  150 K. 

The region  was observed in the position switching mode using the OTF technique with a space between dumps in the scanning direction of 9\arcsec. The rms noise of a single spectrum in the OTF mode was 0.35 K.  The off-source position free of  CO emission was located at \radec\ = (16$^h$36$^m$40.56$^s$, $-$42\gra 3\arcmin 40\farcs 6). Calibration was performed using  Mars and  X-TrA sources. Pointing was done twice during observations using  X-TrA, o-Ceti and VY-CMa. The intensity calibration has an uncertainty of  10\%. The \cdo\ line at 219.560357 GHz was also observed, although the emission is very low and it was not used in the analysis.

\begin{figure}
\centering
\includegraphics[width=250pt]{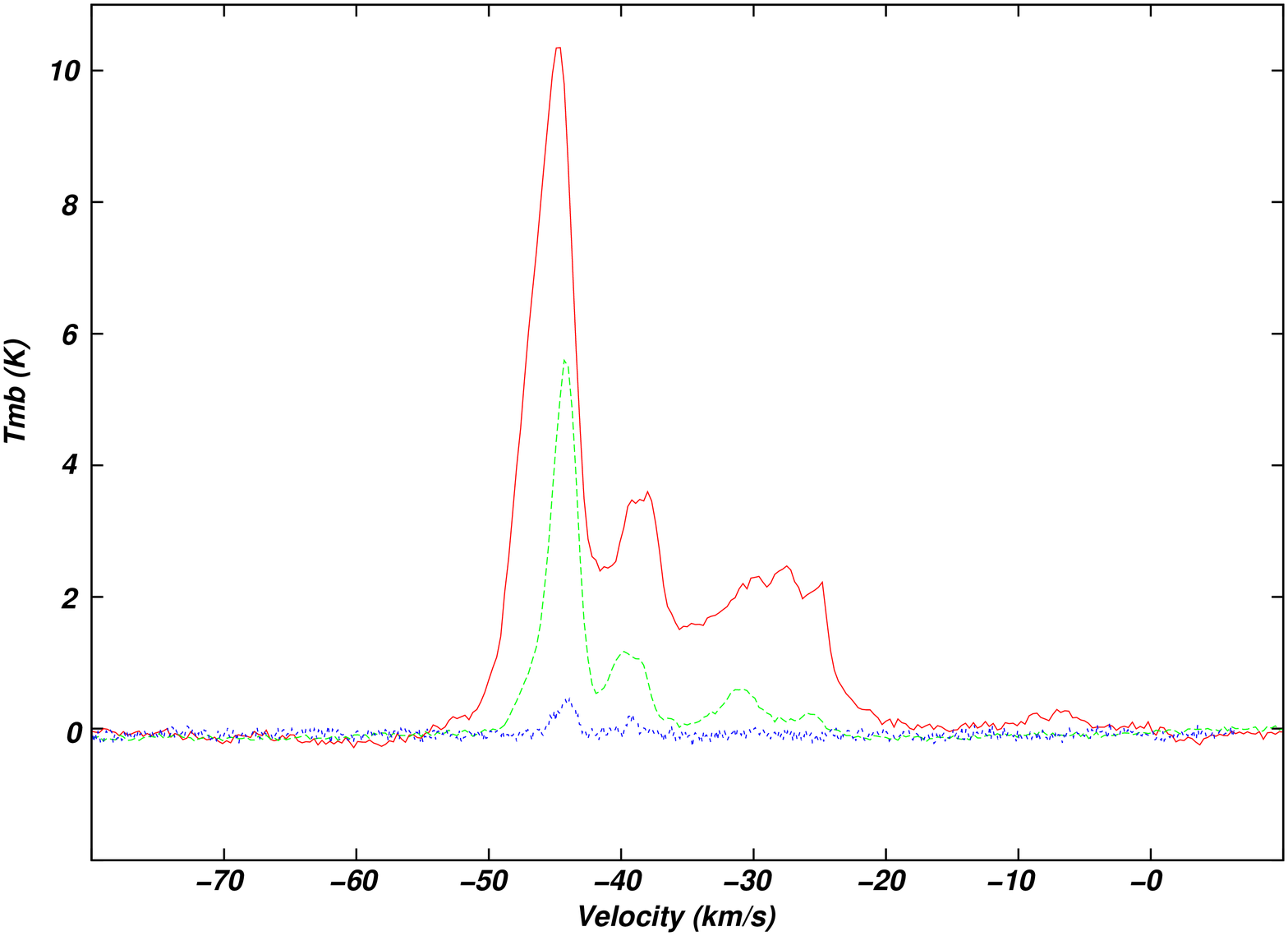}
\caption{ $^{12}$CO(2-1) (red line), $^{13}$CO(2-1) (green line), and \cdo\ (blue line) averaged molecular line spectra obtained toward  S\,21.  }
\label{perfiles}
\end{figure}
\begin{figure}
  \centering
  \includegraphics[width=9cm]{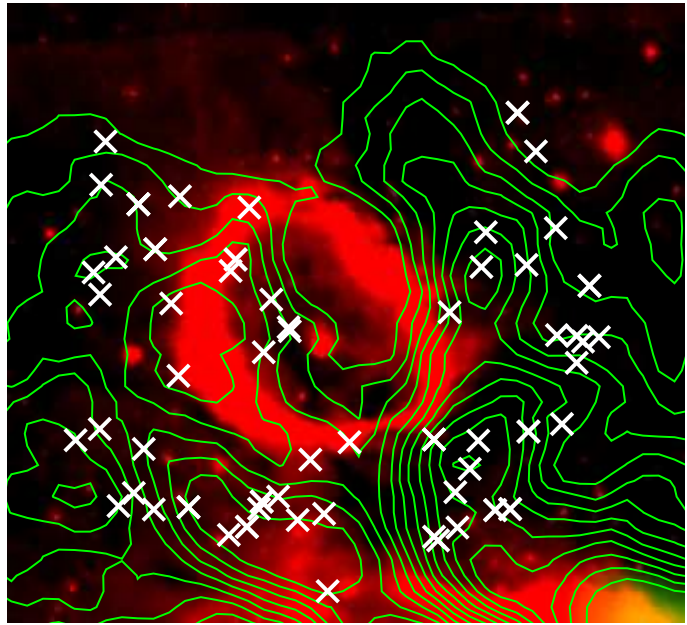}
 \caption{Overlay of the 8 $\mu$m emission (colorscale) and the $^{13}$CO emission (in contours) integrated in the velocity range from $-$45.8 to $-$42.6 \kms. Contour levels go from 3 K ($\sim$ 45 {\it rms}) to 6.9 K in steps  of 0.3 K and from 6.9 in steps of 1 K. The white crosses show the position of candidate YSOs identified in the VVV database projected onto the molecular shell (see Sect. 7). 
}
  \label{shell}
\end{figure}
The spectra were reduced using the Continuum and Line Analysis Single-dish Software (CLASS) of the Grenoble Image and Line Data Analysis Software (GILDAS) working group\footnote{http://www.iram.fr/IRAMFR/PDB/class/class.html}. A linear baseline fitting was applied to the data.  The observed line intensities are expressed as main-beam brightness temperatures $T_{\rm mb}$, by dividing the antenna temperature $T_{\rm A}$  by the main-beam efficiency $\eta_{\rm mb}$, equal to 0.72. The Astronomical Image Processing System (AIPS) package and CLASS software were used to perform the analysis. The final molecular data were smoothed to 0.3 \kms, with a final rms noise of 0.2 K.
 
\subsection{Archival dust continuum data}

\subsubsection{Herschel data}

The archival data comes from the Hi-GAL key program (Hi-GAL:{\em Herschel} Infrared GALactic plane survey, Molinari \etal 2010), OBSIDs: 1342204094 and 1342204095). These data include PACS images at 70 and 160 $\mu$m (Poglitsch \etal 2010) and SPIRE  images at 250, 350, and 500\,$\mu$m (Griffin \etal 2010).The data were re-reduced using the {\em Herschel} Interactive Processing Environment (HIPE v12\footnote{{\em HIPE} is a joint development by the Herschel Science Ground Segment Consortium, consisting of ESA, the NASA Herschel Science Center, and the HIFI, PACS and SPIRE consortia members, see http://herschel.esac.esa.int/HerschelPeople.shtml.}, Ott 2010) as described in Cappa \etal (2016). The angular resolutions of the final dust continuum images spans from 8\arcsec\ to 35\farcs 2 for 70\,$\mu$m to 500\,$\mu$m, respectively. 

\subsection{Complementary data}

We use archival images of ATLASGAL at 870 $\mu$m (345 GHz)  (Schuller \etal 2009). This survey has an rms noise in the range 0.05 - 0.07 \jyb. The calibration uncertainty in the final maps is about of 15$\%$. The {\it Large APEX BOlometer CAmera} (LABOCA) used for these observations, is a 295-pixel bolometer array developed by the Max-Planck-Institut fur Radioastronomie (Siringo \etal 2007). The beam size at 870 $\mu$m is 19\farcs2. 

Also, Spitzer images at 3.6, 4.5, and 8.0 $\mu$m from the Galactic Legacy Infrared Mid-Plane Survey Extraordinaire (GLIMPSE; Benjamin \etal 2003), and at  24 $\mu$m from the MIPS Inner Galactic Plane Survey (MIPSGAL; Carey \etal 2005) were used.

Radio continuum data from the Sydney University Molonglo Sky Survey (SUMSS\footnote{http://www.astrop.physics.usyd.edu.au/cgi-bin/postage.pl}, Bock \etal 1999) at 843 MHz with a resolution of 43\arcsec \x 43\arcsec\ csc(Decl.) and an rms noise level of ~1 \mjyb\ and at 1.4 GHz from the Southern Galactic Plane Survey (beam size = 1\farcm 7, Haverkorn \etal 2006) were used. 

\subsection{Search for young stellar objects}

To investigate the existence of candidates to YSOs projected onto the region we used infrared point sources from the Vista Variables in the Via Lactea ESO Public Survey (VVV, ESO programme ID 179.B-2002; Minniti et al. 2010; Saito et al. 2012), and the Spitzer (Fazio \etal 2004) and Wide-field Infrared Survey Explorer (WISE; Wright \etal 2010) point source catalogues.

\section{Molecular gas linked to the bubble}

\subsection{Molecular gas distribution}

\begin{figure*}
\centering
\includegraphics[width=13cm]{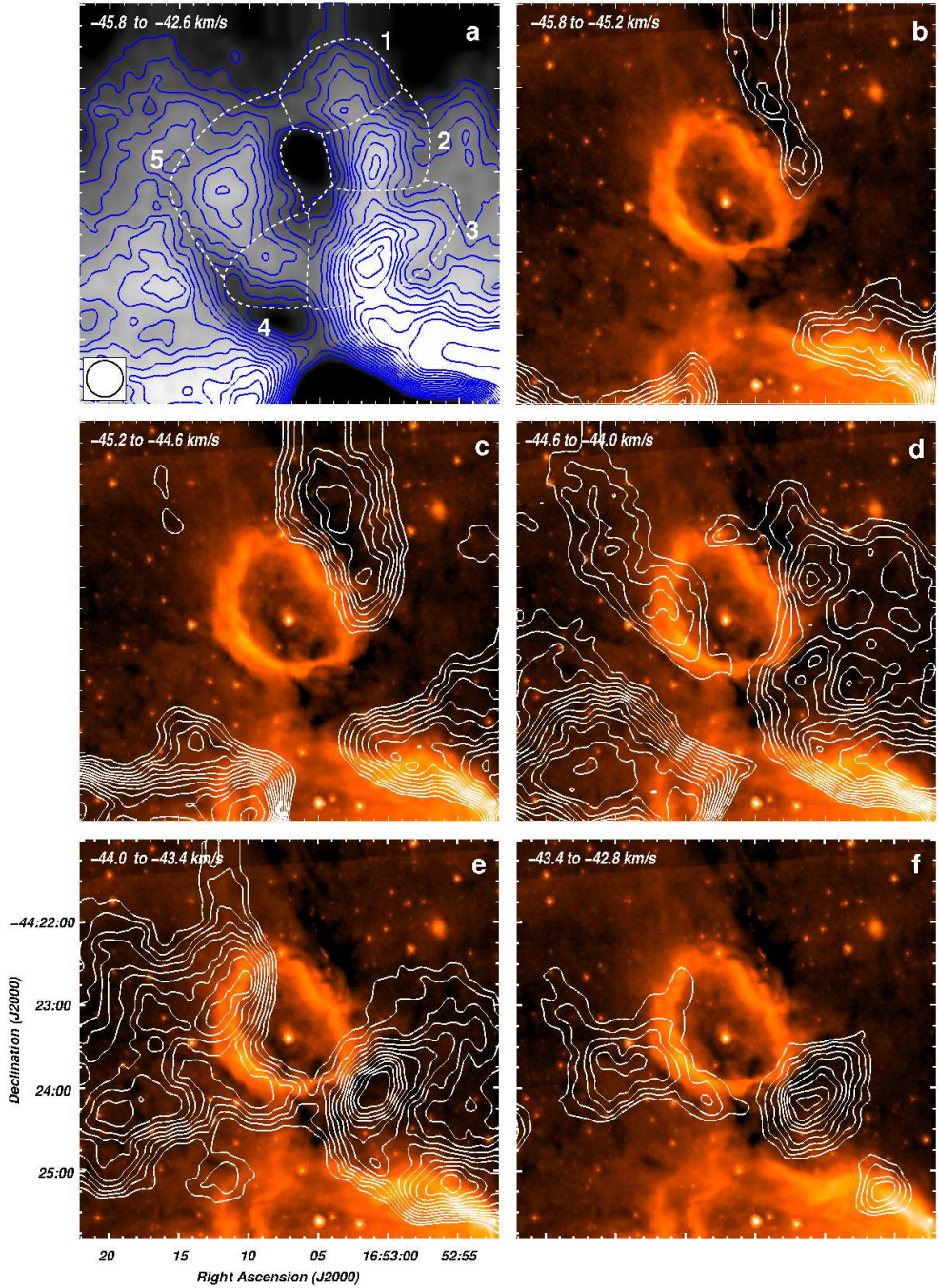}
\caption{{\it Panel  a}: $^{13}$CO emission in the  velocity range from $-$45.8 to $-$42.6 \kms. Contour levels go from 3 K ($\sim$ 45 {\it rms}) to 6.9 K in steps  of 0.3 K and from 6.9 K in steps of 1 K. White dotted lines indicate the limits chosen to define the clumps (see text). The beam size is indicated in the lower left corner of this panel.  {\it Panels  b to  f}: Channel maps of $^{13}$CO in intervals of  0.6 \kms\ (white contours) superimposed on the GLIMPSE 8.0 $\mu$m red colorscale. Contours go from 4.5 K ($\sim$ 32 K {\it rms}) to 10 K in steps of 0.5 K, and from 10 K in steps of 1 K.    }
\label{mosaico}
\end{figure*}

In Fig. 2 we show the $^{12}$CO(2-1), $^{13}$CO(2-1), and \cdo\  spectra averaged in a region $\sim$ 4$'$ $\times$ 4$'$ around the central position of S\,21. The  bulk of the molecular emission  appears concentrated between  $-$55 \kms\ and $-$20 \kms, with three components peaking approximately at $-$45 \kms, $-$38 \kms, and $-$30 \kms. The spatial distribution of these components reveals that only the molecular component peaking at $-$45 \kms\ shows a morphological correspondence with S\,21. The emission distribution in the velocity interval from $-$45.8 to $-$42.6 \kms\ shows a  bubble fully inmersed in its parental molecular cloud.  CO  shells have been reported for many other IRDBs (see for example Arce \etal 2011). Figure~3 displays the integrated $^{13}$CO(2-1) emission distribution in the mentioned velocity range in contours overlayed onto the emission at 8 $\mu$m. A clumpy circular structure coincident with the infrared bubble with a minimum in the emission projected onto its center can be discerned. The mean radius of the molecular shell is 1\farcm 3. The molecular correspondence between the IR emission at 8 $\mu$m and the molecular emission within the quoted velocity interval indicates that the shell is the molecular counterpart of the IR bubble, and that they are physically associated.

\begin{table*}
\caption{Parameters of the  molecular clumps.    }
\begin{center}
{\scriptsize 
\begin{tabular}{cccccccccc}
\hline
 Clump & R.A.(J2000)   & Decl.(J2000) & $T_{peak}^{12}$ & $\Delta{\rm v^{12}}$ & $T_{peak}^{13}$  & $\Delta{\rm v^{13}}$ & $T_{\rm exc}$   &  $\tau^{13}$ & $\tau^{12}$\\
  & (h m s)  & (\gra\ \arcmin\ \arcsec)  & (K)  & (\kms)  & (K) & (\kms) & (K) \\ 
\hline
\hline
1   & 16:53:03 & -44:22:16.3 & 11.5 & 3.6 &  7.5 & 1.6 & 16.6 & 0.96 & 34.3\\
2   & 16:53:00 & -44:22:57.3 &  8.5 & 2.5 &  7.3 & 2.1 & 13.5 & 1.83 & 79.5\\
3   & 16:53:01 & -44:22:09.2 & 13.0 & 4.3 & 10.2 & 1.7 & 18.1 & 1.50 & 33.5 \\
4   & 16:53:08 & -44:22:01.3 &  8.8 & 3.8 &  6.5 & 1.8 & 13.1 & 1.28 & 33.3\\
5   & 16:53:12 & -44:23:13.7 & 10.5 & 4.0 &  7.8 & 1.9 & 15.6 & 1.30 & 33.1\\
\hline
\end{tabular}
\label{observadas}
}
\end{center}
\end{table*}

\begin{table*}
\caption{Properties derived for the molecular clumps.    }
\begin{center}
{\scriptsize 
\begin{tabular}{cccccccccc}
\hline
Clump &   $\Delta {\rm v}$  &  $T_{\rm mean-mb}^{13}$ &  $\Delta {\rm v}T_{\rm mean-mb}^{13}$ &  $N(^{13}\rm CO)$   &  R$_{eff}$ & $M({\rm H_2})$ & $n_{H2}$ & $M_{VIR}$ &  $\frac{M_{VIR}}{M({\rm H_2})}$\\
  &  (\kms)     &  (K) & (K \kms) & (10$^{16}$ cm$^{-2}$)   &  (pc) &  (\msun) & (10$^3$ \cmtres)& (\msun) \\ 
\hline
\hline
1   &  5.0 & 3.75 & 12.0 & 2.4 & 0.65 & 350$\pm$105 & 4.6 & 209-316 & 0.5-1.3\\   
2   &  5.4 & 4.52 & 14.4 & 4.0 & 0.68 & 650$\pm$195 & 7.5 & 377-519 & 0.4-1.1\\   
3   &  3.7 & 5.25 & 16.8 & 4.2 & 0.77 & 880$\pm$260 & 7.0& 280-422 & 0.2-0.7\\   
4   &  4.8 & 3.85 & 12.3 & 2.8 & 0.64 & 400$\pm$120 & 5.5 & 261-393 & 0.3-1.4\\   
5   &  4.6 & 4.39 & 14.0 & 3.2 & 0.75 & 620$\pm$185 & 5.3 & 341-514 & 0.4-1.2\\   
\hline
\end{tabular}
\label{propiedades}
}
\end{center}
\end{table*}

Using clumpfind we have identified  several condensations in the molecular structure around S\,21, which will be referred as \lq\lq clumps\rq\rq\ (Blitz 1993; Williams \etal 2000).   The location  of the clumps, labeled from 1 to 5, are depicted in Fig.~4a. We were left only with clumps adjacent to the bubble, which were very likely formed in the collected layers of the molecular gas.  The  coordinates and peak temperatures of the clumps are listed in Table~1.   

In order to unambiguously ascertain the relationship of each clump with the nebula and to provide a better visual display, in panels {\it b} to {\it f} of Fig.~4 we show the channel maps of $^{13}$CO(2-1) in velocity intervals of 0.6 \kms, overlaid onto the GLIMPSE 8.0 $\mu$m emission image. Between $-$45.8 to $-$45.2 \kms\ (Fig.~4b) clumps 1 and 2 become noticeable, bordering the bubble from its north-western side. In this velocity range, two bright molecular features can be also detected to the south of S\,21, unconnected to this bubble. Clumps 1 and 2 achieve their maximum brightness temperature in the velocity interval from $-$45.2 to $-$44.6 \kms\ (Fig.~4c). Clump 2 is still detected between  $-$44.6 to $-$44.0 \kms\ (Fig.~4d), where clump 3 becomes first  noticeable. In this velocity interval, clump 5 appears as an extension of an elongated  feature that borders the eastern side of S\,21. From  $-$44.0 to $-$43.4 \kms\ (Fig.~4e) clumps 3, 4, and 5 attain their peak temperature. They  surround the nebula from south to east (clump 3) and  from south to north (clumps 4 and 5). Finally, between $-$43.4 to $-$42.8 \kms\ (Fig.~4f), clumps 4 and 5  are barely detected but clump 3 is very  bright. Clump 3 dissapears at $\sim$ $-$42.1 \kms\ (not shown here).  At a distance of 3.7 kpc, the radius of the molecular shell is $\sim$ 1.4 pc.

The eastern section of the molecular shell (clumps 4 and 5) coincides with the photodissociation region (PDR), while the western section (clumps 1, 2, and 3) is projected onto regions of low emission at 8 $\mu$m close to the PDR.
 
\subsection{Physical parameters of the molecular gas}

We have estimated some properties for the identified clumps, which are presented in Tables~1 and 2. Assuming that all rotational levels are thermalized with the same excitation temperature (LTE conditions) and that the emission in the $^{12}$CO(2-1) line is optically thick, we derived the  excitation temperature $T_{\rm exc}$ (Column 8, Table~1) from the emission in the $^{12}$CO(2-1) line  using 
\begin{equation}
T_{\rm peak}^{12} = T_{12}^*\left[\left(e^{\frac{T_{12}^*} {T_{\rm exc}}}-1\right)^{-1}
-\left(e^{\frac{T_{12}^*} {T_{\rm bg}}}-1\right)^{-1}\right]
\label{texc12co}
,\end{equation}
where $T_{\rm 12}^*$ = $h \nu_{\rm 12} / k$, being $\nu_{\rm 12}$ the frequency of the $^{12}$CO(2-1) line,  and $T_{\rm bg} = 2.7$ K. 
To obtain the peak main-beam brightness-temperature in the $^{12}$CO(2-1) line $(T_{\rm peak}^{12})$ (column 4, Table~1) we used the spectrum toward the position of maximum emission of the clump.

The optical depth $\tau^{13}$ (Column 9, Table~1)  was obtained from the $^{13}$CO(2-1) line  by assuming that the excitation temperature is the same for $^{12}$CO(2-1) and $^{13}$CO(2-1) emission lines using the expression 
\begin{equation}
\tau^{13}=-{\rm ln}\left[1-\frac{T_{\rm peak} {\rm ^{13}}}{T_{13}^*}
\left[\left(e^{\frac{T_{13}^*} {T_{\rm exc}}}-1
\right)^{-1}-\left(e^{\frac{T_{13}^*} {T_{\rm bg}}}-1\right)^{-1}\right]^{-1}\right],
\label{tau13co}
\end{equation}
where $T_{\rm 13}^*$ = $h \nu_{\rm 13} / k$, being $\nu_{\rm 13}$ the frequency of the $^{13}$CO(2-1) line. We also estimated the optical depth of the $^{12}$CO(2-1) line (Column 10, Table~1)  from the $^{13}$CO(2-1) line with  
\begin{equation}
\quad \tau^{12} =\  \left[\frac{\nu^{13}}{\nu^{12}}  \right]^2\ \times  \left[\frac{\Delta {\rm v}^{13}} {\Delta   {\rm v}^{12}} \right]\ \times  \left[\frac{^{12}{\rm CO}}{^{13}{\rm CO}} \right]\   \tau^{13},
\label{tau12}
\end{equation}
where   $^{12}$CO/$^{13}$CO is the isotopic ratio (assumed to be $\sim$ 62; Langer \& Penzias 1993);  $\Delta{\rm v}^{13}$ and $\Delta{\rm v}^{12}$ are the {\it full width at half  maximum} (FWHM) of the spectra  of the $^{13}$CO and $^{12}$CO lines, respectively. These values are indicated in columns 5 and 7 of Table~1.  

In LTE,  the $^{13}$CO column density (Column 5, Table~2)  can be estimated from the $^{13}$CO(2-1) line data following
\begin{equation}
N({\rm^{13}CO})=2.4\times 10^{14} \left[\frac{e^{  \frac{T_{13}^*}{T_{\rm exc}}  }}{1 -e^{-  \frac{T_{13}^*}{T_{\rm exc}}   } }\right]  T_{\rm exc} \int \tau^{13}\   dv \ \  \textrm{ (cm$^{-2}$)}
\label{n13co}
.\end{equation}
The integral of Eq. 4 can be approximated by
\begin{equation}
T_{exc} \int{\tau^{13} dv \approx\ \frac{\tau^{13}}{1-e^{(-\tau^{13})}}
\int{T_{\rm mb}}}\ \ d{\rm v}.
\label{integral}
\end{equation}
This approximation helps to eliminate to some extend optical depth effects and is good within 15\% for $\tau <$  2 (Rohlfs \& Wilson 2004). Considering that $\tau^{\rm 13} <$ 2 for the clumps, this approximation is appropriate for our region. The integral was evaluated as  $\Delta {\rm v}T_{\rm mean-mb}$ (with $T_{\rm mean-mb}$ equal to the average $T_{\rm mb}$  within the area of the clump) and is listed in column 4, Table~2.  Then, the total hydrogen mass (Column 7, Table~2) can be calculated using
\begin{equation}\label{eq:masa}
{\rm M(H_2)}\ =\ (m_{\rm sun})^{-1}\ \mu\ m_H\  A \ {\it N}(\rm H_2)\ {\it d}^2 \quad \quad \quad \textrm{($M_\odot$)}
,\end{equation}
where $m_{\rm sun}$ is the solar mass ($\sim$2$\times$10$^{33}$ g), $\mu$ is the mean molecular weight, which is assumed to be equal to 2.76 after allowing for a relative helium abundance of 25\% by mass Allen (1973), m$_{\rm H}$ is the hydrogen atom mass ($\sim$1.67$\times$10$^{-24}$ g),  $A$ is the solid angle of the  $^{13}$CO emission (included in Table~2 as the effective radius $R_{eff}$ = $\sqrt{A/\pi}$, Column 6), and $d$ is the adopted distance expressed in cm. To obtain the masses, we adopted an abundance \hbox{$N(\rm H_2)$ / $N(^{13} {\rm CO})$} = \hbox{5$\times$10$^{5}$} (Dickman 1978). Uncertainties in molecular masses  are about 30\%, while they are about 50\%\ in ambient densities, and originate mainly in distance uncertainties.  

Mean volume densities of the clumps are in the range (4.5-7)\x 10$^3$ \cmtres\ (Column 8 of Table~2). We estimate the original mean volume  ambient density in the region of the bubble by assuming a uniform gas distribution before the ring was formed. This density was obtained by distributing the total shell mass within a sphere with the outer radius of the shell (1.7 pc),  and amounts to 2.1\x 10$^3$ \cmtres. This value indicates that the bubble is evolving in a high density interstellar medium. 

\begin{figure*}
  \centering
  \includegraphics[width=11cm]{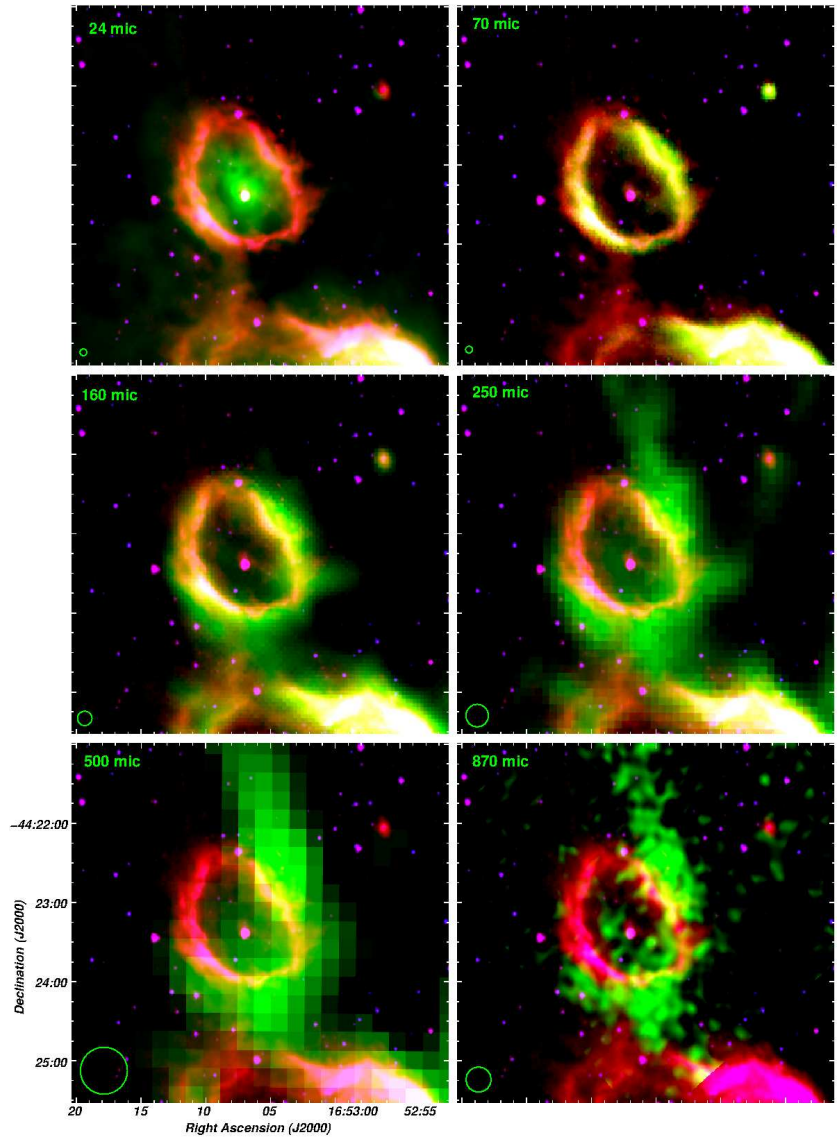}
\caption{Composite images showing the emission in the near-, mid-, and far-infrared. All the images display the emissions at 8 $\mu$m (in red) and at 3.6 $\mu$m (in blue) from IRAC-GLIMPSE. The emissions in green correspond to different IR wavelengths indicated in the upper left corner of each image. The angular resolution of the IR images is shown in the lower left corner. }
  \label{dust}
\end{figure*}

\section{The interstellar dust linked to the bubble}

\subsection{Dust distribution}

Dust associated with the bubble can be analyzed from the distribution of the emission at different wavelengths in the infrared.  

Each panel of Fig.~5  shows a composite image  of the emissions at 8 $\mu$m (in red) and 3.6 $\mu$m (in blue) from IRAC-GLIMPSE, and the emissions at 24, 70, 160, 250, 500, and 870 $\mu$m (in green) from MIPS, {\it Herschel}-PACS and -SPIRE, and ATLASGAL. 

The emission at 24 $\mu$m  is seen projected onto the inner part of the bubble, with its maximum near the central point source. The emission at 70 $\mu$m coincides fairly well with that at 8 $\mu$m. The emission differs significantly at larger wavelengths. At 160 $\mu$m it encircles externally the bubble completely and appears more extended than at 70 $\mu$m. At 250 and 500 $\mu$m, the emission is brighter than at lower wavelengths and extends toward the western and northern parts of the bubble. The SPIRE emission distribution at 350 $\mu$m (not shown here) is similar to that at 250 and 500 $\mu$m. Similarly to the case of S\,24, cold dust emission detected at 870 $\mu$m resembles that at 250 and 500 $\mu$m, although the emission at 250 and 500 $\mu$m, which also shows the distribution of cold dust, appears more extended. We have to bear in mind that large scale dust continuum emission might be filtered out in LABOCA bolometric observations. 

Two facts can be concluded from these images. On one hand the spatial emission distribution in the mid- and far-IR seems to show a gradient in dust temperature, with lower values in the outer regions of the bubble where continuum emission at larger wavelengths dominates. Indeed,  emission at $\lambda > $ 160 $\mu$m is present well outside the PDR. On the other hand, the emission at 24 $\mu$m inside S\,21 is indicative of the presence of exciting sources. Both statements  will be analyzed in some detail in the next sections.

A comparison with the molecular gas distribution around the bubble (see Fig.~6) shows that clumps 1 to 4 partially coincide with the cold dust counterpart identified in the {\it Herschel}-SPIRE and in the LABOCA images. 

The observed dust continuum emission at different wavelengths in the IR coincides with previous findings toward other IR dust bubbles.

The molecular emission from the $^{12}$CO(2-1) line may contribute to the thermal emission at 870 $\mu$m. In our case, this contribution is less than  about 1\% of the emission at 870 $\mu$m, and consequently, within calibration uncertainties. The other process that contribute to the emission at  870 $\mu$m is the free-free emission from ionized gas. Their contribution, bearing in mind the flux density at 1.4 GHz (see Sect. 5), is less than 0.1\% of the emission at 870 $\mu$m, and again, within calibration uncertainties. 

\subsection{Dust temperatures and mass}

Dust temperatures T$_{\rm dust}$ in the environs of the S\,21 region can be obtained using the SPIRE images at  250 $\mu$m and 350 $\mu$m.
To perform this we convolved the image at 250 $\mu$m down to the angular resolution at 350 $\mu$m and assumed that the emission is optically thin. The {\it color-temperature} map  was constructed as the inverse function of the ratio map of Herschel 250 $\mu$m and 350\,$\mu$m color-and-background-corrected maps, i.e., T$_{\rm dust} = f(T)^{-1}$, where $f(T)$ is:
\begin{equation}
f(T) = \frac{S_{250}}{S_{350}} = \frac{B_\nu(250\,\mu m,{\rm T})}{B_\nu(350\,\mu m,{\rm T})} \left( \frac{250}{350} \right) ^{\beta_{\rm d}}
\end{equation}
\noindent In this expresion $S_{250}$ and $S_{350}$ are the flux densities in \jyb, $B_\nu(\nu,{\rm T})$ is the blackbody Planck function and $\beta_{d}$, the spectral index of the thermal dust emission. The pixel-to-pixel temperature was calculated assuming $\beta_{\rm d} = 2$. This is a typical value adopted for irradiated regions.

The dust temperature map is shown in Fig.~7. The highest dust temperatures (33 K) are present at the NE extreme of the 8 $\mu$m bubble. Values in the range 24-33 K coincide with the eastern section of the bubble, while lower values (21 K) were obtained for the western side. Low dust temperatures coincide with molecular gas and with regions  with faint emission at 8 $\mu$m. Watson \etal (2010) derived dust temperatures for the interior of S\,21 based on images from MIPS at 24 $\mu$m (whose emission is detected inside the bubble) and 70 $\mu$m (detected up to the border of the bubble).  Our estimates, based on images in the far IR, sample colder dust present in the outkirts of the bubble.

\begin{figure}
  \centering
  \includegraphics[width=8cm]{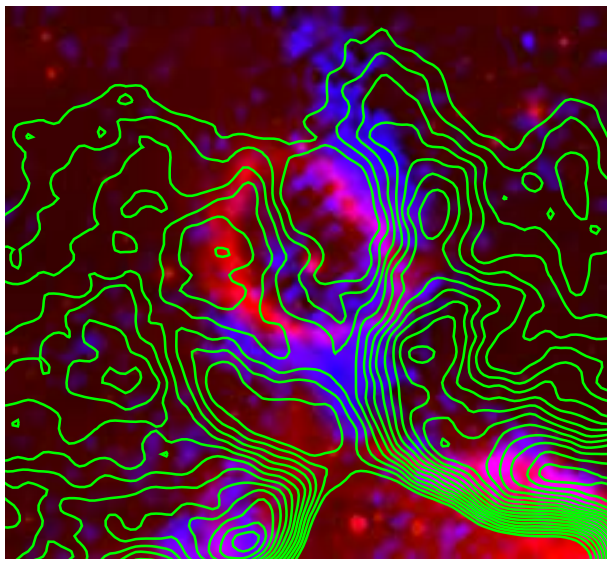}
 \caption{ Composite image showing the emission at 8 $\mu$m (in red) and at 870 $\mu$m (in blue), and the same  $^{13}$CO contours  of Fig. 3.}
  \label{co-dust}
\end{figure}

\begin{figure}
  \centering
  \includegraphics[width=8cm]{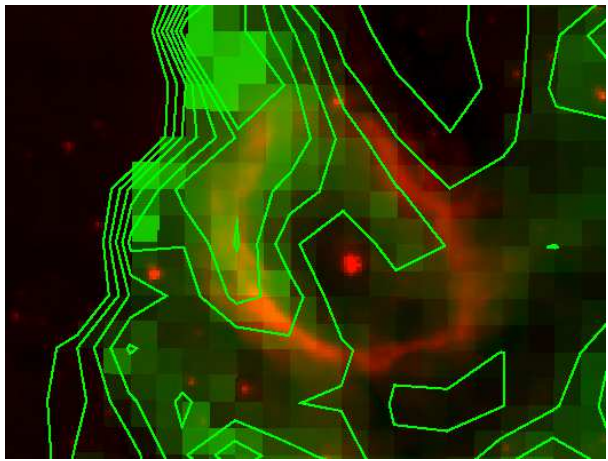}
 \caption{Dust temperature map derived from the Herschel emission at 250 and 350 $\mu$m superimposed onto the image at 8 $\mu$m. Blue color scale goes from 15 to 33 K.  Brighter  blue regions indicate higher dust temperatures. Contour levels correspond to 15 to 30 K, in steps of 3 K.}
  \label{tempera}
\end{figure}

Dust masses can be estimated from the expression (Hildebrand 1983)
\begin{equation}
{\rm M}_{\rm dust} = \frac{S_{\rm 870} \ d^{2}}{\kappa_{\rm 870} \ B_{870}(T_{\rm dust})}
\end{equation}
\noindent where $S_{\rm 870}$ is the flux density at 870 $\mu$m, $d$ = 3.7$\pm$0.5 kpc, $\kappa_{\rm 870}$ = 1.0 cm$^{2}$/gr is the dust opacity per unit mass (Ossenkopf \& Henning 1994), and $B_{\rm 870}(T_{\rm dust})$ is the Planck function for a temperature T$_{\rm dust}$. 

The flux density $S_{\rm 870}$ obtained by integrating the emission over the observed emitting area linked to S\,21 at this wavelength (see Fig.~5f) amounts to 1.9$\pm$0.4 Jy. Adopting a mean value $T_{\rm dust}$ = 30 K for S\,21, a dust mass $M_{dust}$ = 1.53$\pm$0.90 \msun\ can be estimated. For gas-to-dust ratios in the range 100-186 (Beuther \etal 2011), the gas mass amounts to 153-285 \msun. 

\section{The ionized gas}

\begin{figure}
  \centering  
\includegraphics[width=9cm]{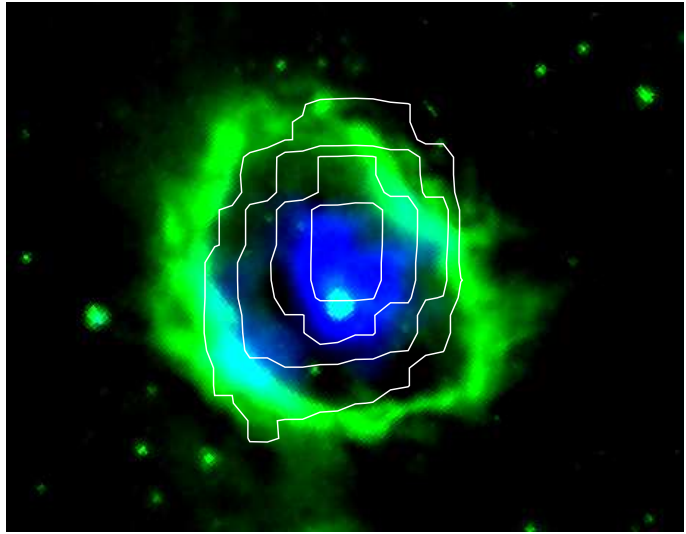}
 \caption{ Overlay of the radio continuum emission at 843 MHz (white contours), the IRAC emission at 8 $\mu$m (in green), and the MIPSGAL emission at 24 $\mu$m (in blue). Contours correspond to 10, 15, 20, and 25 \mjyb.}
  \label{cont-radio}
\end{figure}

Figure~\ref{cont-radio} shows an overlay of the SUMSS image at 843 MHz (in white contours) and the emissions at 8 $\mu$m and 24 $\mu$m (in colorscale). The 843 MHz image (synthesized beam = 43\arcsec$\times$62\arcsec) shows a radio source coincident with S\,21 catalogued by Murphy \etal(2007) with a size of 75\farcs 7$\times$69\farcs 6  and a flux density $S_{\rm 0.843}$ = 49.2 mJy. The elongated shape of the source is due to the synthesized beam of the data. The source is also detected at 1.4 GHz (SGPS, Haverkorn \etal 2006) with an estimated flux density of  $S_{1.4}$ = 52 mJy. With these values, the derived spectral index $\alpha \simeq$  +0.1 ($S_{\nu} \propto \nu^\alpha$). Within uncertainties, this value  is consistent with thermal emission of an \hii\ region optically thick  at 843 MHz.

Thus, these results indicate the presence of ionized gas, and consequently, the existence of at least one exciting source, compatible with warm dust inside the bubble as shown by the 24 $\mu$m image. The emission at 8 $\mu$m due to PAHs reveals a PDR bordering the ionized region and the presence of molecular gas in its exterior. The radius of the ionized region inside the 8 $\mu$m dust bubble is 38\arcsec\  or 0.7 pc at 3.7 kpc.  The derived spectral index confirms the classification of S\,21 as \hii\ region by Anderson \etal(2015). The characteristics of the region, visible in the IR and in the radio continuum, as well as the large mean original ambient density of 2100 \cmtres\ (see Sect. 3.2) suggest a compact \hii\ region (Urquhart et al. 2013). 

To produce the observed flux at 1.4 GHz we requiere an UV photon flux that can be obtained from Matsakis \etal (1976)
\begin{equation}
N_{Ly} = 7.5 \times 10^{43} S_{1.4} \nu^{0.1} d^2 T_e^{-0.45} \  \  s^{-1}
\end{equation}
where $S_{1.4}$ is in units of mJy, $\nu$ in GHz, $d$ in kpc, and  the electron temperature $T_e$ in 10$^4$ K. Considering the galactic electron temperatures for \hii\ regions (Quireza \etal 2006), we assumed  $T_e$ = 7000 K. The required UV flux amounts to  6.5$\times$10$^{46}$ s$^{-1}$. This ionizing flux is underestimated since part of the stellar UV photons are used to heat the dust. Considering that half of the stellar photons are absorbed by dust (Inoue 2001), the UV photon flux would be 1.3$\times$10$^{47}$ s$^{-1}$. According to Martins \etal (2005), this value indicates that the ionization of the gas could be produced by at least a  O9.5V or earlier type star.  

\section{Search for an exciting star}

Identifying the exciting star of this \hii\ region is not an easy task.  The extinction towards the inner part of the IR bubble can help to identity this star since we expect a similar extinction. This extinction can   be deduced from the expresion  by Bohlin \etal (1978)
\begin{equation}
N(HI) + 2N(H_2) = 5.8 \times 10^{21} E(B-V)
\end{equation}
Considering only the $H_2$ column density toward the IR source and taking into account  $N(H_2)$ =  2.9$\times $10$^{21}$ \cmdos\ (estimated from the  {\bf $^{13}$CO} emission), we calculate a visual absorption of 30 mag. 

To search for exciting stars the color-magnitude diagrams were build using the VVV DR4 catalog (Fig.~8). The foreground Main Sequence stars are situated around $J - K_s$ =  0.7 mag, and the mean reddening is estimated as $E(J-K_s)$ = 0.5 mag, some field red giants can be also identified.  The 76 extremely red stars ($(J-K_s) >$ 4 mag) are projected in the field (see also the lower panel of Fig.~9), seven of them are projected within the inner radius of the bubble and could be candidates of the exciting source. However, all of them are too faint and a spectroscopic follow up is necessary to reveal their nature.

As pointed out in Sect. 1, \lafuente, located close to the center of the IR bubble, is detected both in the optical and at IR wavelengths, appearing saturated in the VVV images. The analysis of the emission of this IR source using VOSA\footnote{http://www.svo2.cab.inta-csic.es/theory/vosa. } and TLUSTY\footnote{http://www.nova.astro.umd.edu} tools and all the available detections at different wavelengths (taking into account stellar atmosphere modelling for O and B stars) suggests that the star might be a O8V of a B0I star. From the comparison of the observed colors taken from the 2MASS catalogue (source 2MASSJ 16530711-4423239) with the intrinsic magnitudes and colors taken from Martins \& Plez (2006) for O8V stars and from Bibby \etal (2008) for B0I stars, we estimate visual absorptions and distances of 21 mag and 0.58$\pm$0.9 kpc for an O8V type star and 20.9 mag and 1.7$\pm$0.8  kpc for a B0I type star, using the standard reddening low and Bessel \etal (1998) transformations.  The calculated distance is not comparable with the kinematical distance of 3.7 kpc to the complex. Thus, G341.3553-00.2885 is most probably a foreground object. Spectroscopic data are necessary to verify this suggestion and unambiguously identify the exciting star of this bubble, which are beyond the scope of this paper.

\begin{figure}
\includegraphics[width=9cm]{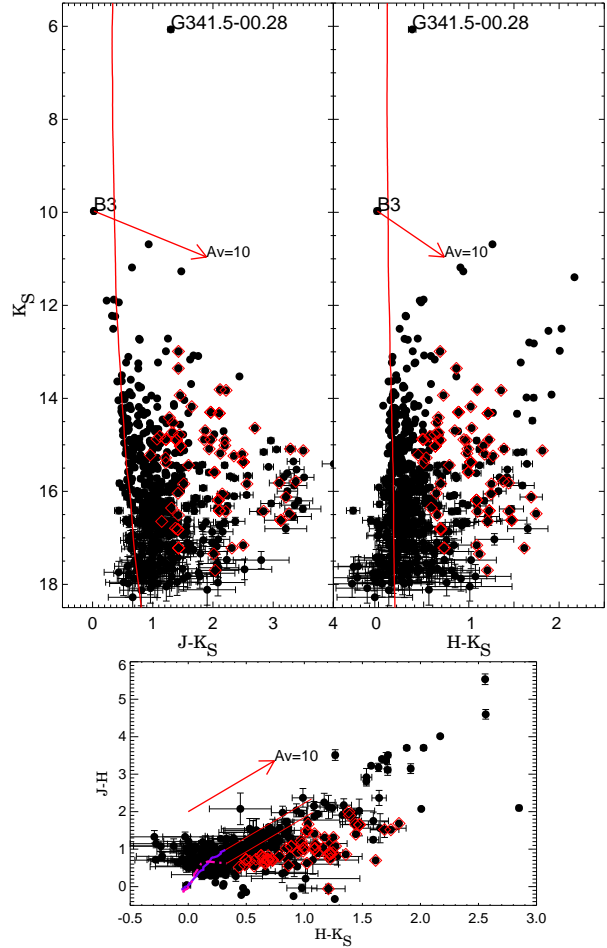}
\vspace{0.1cm}
\caption{{\it Upper panel.}  $(J-K_s)$ vs $K_s$ and $(H-K_s)$ vs $K_s$ VVV color magnitude diagrams within a radius of 1\farcm5  centered on S\,21. The vertical red line is the Zero Age Main Sequence, reddened with $E(J-K_s)$ = 0.5 mag (the mean reddening of the field stars).  The candidate YSOs projected onto the molecular shell are overplotted with red squares. The red arrow shows the reddening vector corresponding to Av=10 mag. \lafuente\  and HD\,329056 sources are labeled.
{\it Bottom panel.} The $(H-K_s)$ vs $(J-H)$ color-color diagram.   The continuous and dashed lines represent the sequence of the zero-reddening stars of luminosity classes I (Koornneff \etal1983) and V (Schmidt-Kaler 1982).
  }
\label{vvv}
\end{figure}

\section{Star formation toward S\,21}

As pointed out in Sect. 1, star formation can be favoured in the molecular and cold dust clumps in the environs of \hii\ regions. To help testing if this is the case for the S\,21 bubble, we can analyze the stability of the molecular clumps by comparing virial to LTE masses. Following MacLaren \etal(1988), the virial mass can be obtained as
\begin{equation}  
\frac{M_{\rm vir}}{M_{\odot}} = k_2 \left[\frac{R}{pc}\right] \left[\frac{\Delta {\rm V}^
2}{km s^{-1}}\right]
\end{equation}

\noindent where $R$ and $\Delta$V are the radius of the region and the velocity width measured from the \tco\ emission, and $k_2$ depends on the density distribution in the clump, being 190 or 126 according to $\rho \propto r^{-1}$ or $\rho \propto r^{-2}$, respectively. Virial masses are included in column 9 of Table~\ref{propiedades} for the two density profiles. The ratio  $\frac{M_{\rm vir}}{M(H_2)}$ = $\gamma$ is listed in column 10. 

As classical virial equilibrium analysis establishes, a ratio $\gamma >$ 1 would imply that the clumps could be stable against collapse, while lower values indicate that collapse is possible. The derived ratio for clump 3 is consistent with collapse, while results for the other four clumps are not conclusive.  

Both virial and LTE masses have large uncertainties due to  distance indetermination (30\% for $M(H_2)$ and 15\% for $M_{VIR}$). Minor errors are due to uncertainties in the boundaries of the clumps resulting in errors in their areas and in the LTE masses.  
 
Virial masses are not free of additional errors, since the existence of magnetic fields might overestimate by up to a factor of 2 the derived values (MacLaren \etal 1988). Other source of error is the density profile of the clump, which is unknown. Uncertainties in distance in the LTE mass and those due to the density profile in virial masses were taken into account in the value of $\gamma$.

\begin{figure*}
\includegraphics[width=11cm]{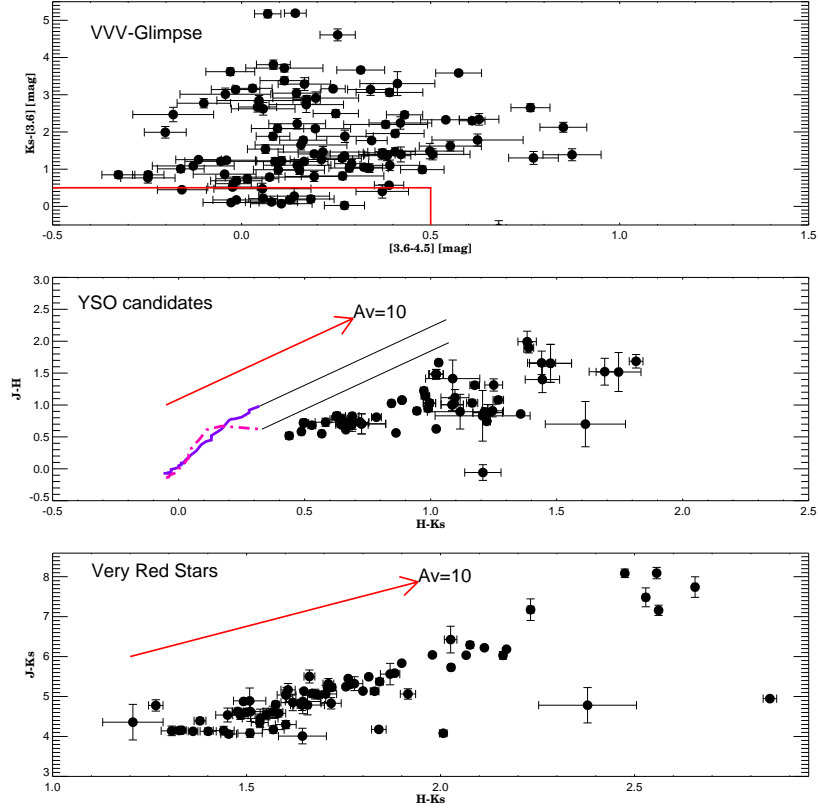}
\caption{{\it Upper panel.} The $K_{\rm S} - [3.6]$,$[3.6] - [4.5]$ color-color plot of stars that are detected in GLIMPSE I. The red dashed lines represent the limits used to select class I and class II YSOs. {\it Middle panel.} The $(J-H)$ vs. $(H-K_{\rm S})$ color-color diagram of the sample. The continuous and dashed lines represent the sequence of the zero-reddening stars of luminosity classes I (Koornneef \etal 1983) and V (Schmidt-Kaler 1982). {\it Lower panel.} The extremely red stars detected in the field.
}
\label{yso_ir_excess}
\end{figure*}

To investigate if YSOs are detected toward the molecular clumps we performed a search for candidates in the available point source catalogues within a region of 1\farcm 5 centered in the bubble. In the region of the bubble and the surrounding molecular shell, no sources with characteristics of YSOs were identified in the Spitzer and WISE databases by applying color criteria (Allen \etal 2004 for Spitzer data; Koenig \etal 2012 for WISE data), mainly because WISE data are not deep enough to measure the fainter stars. 

To select new candidate YSOs in the studied region we used the VVV database and applied photometric criteria. First, from the near-infrared $(J-H)/(H-K)$ color - color diagram we selected all stars which are at least 3$\sigma$ distant from the reddening line that marks the colors of dwarf stars. The list thereby obtained was cross-matched with GLIMPSE measurements. They are shown in Fig.~10. The objects with [$K_{\rm S} - [3.6]]> 0.5$ or [$[3.6] - [4.5]]> 0.5$ magnitudes are considered as a most probable class I and class II YSOs. These limits are set in order to avoid selecting objects that are more likely class III objects or normal stars (dashed red line in Fig.~10). Our final list contains 71 YSO candidates. 

The location of the sources in the color magnitude and color-color diagrams is shown in Fig. 9, while Fig.~3 shows their spatial correlation with the molecular clumps. The presence of candidate YSOs projected onto the molecular shell suggests that star formation has been active recently. However, it is not possible to determine if all candidate YSOs are linked to the molecular shell around S\,21.

The widely known {\it collect-and-collapse} mechanism (C\&C; Elmegreen \& Lada 1977) proposes that the expansion of an ionization front over its parental molecular cloud can trigger the star formation process. The molecular gas may fragment alongside the ionization front as it expands and the fragments may become inestable, giving rise to a new generation of stars. To test whether the C\&C  mechanism  have triggered star formation in the molecular shell around S\,21 we apply the analytical model by Whitworth \etal (1994). For the case of \hii\ regions, the model predicts the age of the \hii\ region at which the fragmentation occurs (the fragmentation time scale), $t_{\rm frag}$, the size of the \hii\ region at that moment, $R_{\rm frag}$, the mass of the fragments, $M_{\rm frag}$, and their separation along the compressed layer, $r_{\rm frag}$. The parameters required to derive these quantities are the UV photon flux of the exciting star, $N_{\rm Ly}$, the ambient density of the surrounding medium into which the \hii\ region is evolving, n$_{0}$, and the isothermal sound speed in the shocked gas, $a_{\rm s}$.

To estimate these parameters we take into account a large range of spectral types, i.e. from O3V to O9.5V stars, with UV fluxes in the range $N_{\rm Ly}$ = (43-0.4)$\times$10$^{48}$ s$^{-1}$ (Martins \etal 2005). Using the mean  $H_{\rm 2}$ ambient density $n_{H2}$ = 2100 \cmtres\ (see Sect. 3.2), and $a_{\rm s}$ = 0.2-0.6 \kms,  we obtained $t_{\rm frag}$ = (1.0-1.5)$\times$10$^6$ yr, $R_{\rm frag}$ = 4.3-3.0 pc, $M_{\rm frag}$ = 20-29 \msun, and, $r_{\rm frag}$ = 0.5-0.3 pc. 

The dynamical age of the \hii\ region can be estimated using the equation (Dyson \& Williams 1997)
\begin{equation}
t_{\rm dyn} \ = \frac{4 R_{\rm S}}{7 c_{\rm s}} \left[ \left(\frac{R}{R_{\rm S}}\right)^{7/4} \ -1\right]
\end{equation}
\noindent where $R_{\rm S}$ is the original Str\"omgren radius, equal to 0.13-0.63 pc for the adopted spectral types, and $c_{\rm s}$ is the sound velocity in the ionized gas. Derived dynamical ages span the range (1.4-33)$\times$10$^4$ yr. We find that the dynamical age is significantly smaller than the fragmentation time scale  $t_{\rm frag}$ for the adopted ambient density, and then, the C\&C process does not seem to be responsible for the triggering of star formation in the envelope. An RDI scenario could be investigated, however evidences of this process (such the presence of pillars) appear to be absent. The  rest of the parameters, $R_{\rm frag}$, $M_{\rm frag}$, and $r_{\rm frag}$, confirm that the \hii\ region is too young to start triggering.

\section{Conclusions}

We performed a multiwavelength study of the IR dust bubble S\,21 using APEX observations of the \dco\ and  \tco\  lines and complementary  images in the near-, mid-, and far-IR from IRAC-Glimpse, MIPSGAL, Herschel, and ATLASGAL.
 
The molecular emission in the $^{12}$CO(2-1) and $^{13}$CO(2-1) lines obtained with the APEX telescope toward the IR dust bubble S\,21 revealed a molecular shell encircling the bubble and partially coincident with the PDR shown by the IRAC emission at 8 $\mu$m. With a mean radius of 1.4 pc, the molecular shell is larger than the  8 $\mu$m bubble. This shell is detected in the velocity interval from --45.8 to --42.6 \kms. The velocity of the shell confirms that S\,21 belongs to the same complex than S\,24. Five clumps were identified in the molecular shell, with radii in the range 0.64-0.75 pc, LTE masses of 350-880 \msun, and volume densities of (4.5-7)$\times$10$^3$ \cmtres. Virial masses for the clumps suggest that at least one of them can collapse. The original ambient density in the region was about 2100 \cmtres.

Complementary  images in the near-, mid-, and far-IR from IRAC-Glimpse, Herschel, and ATLASGAL were used to characterize the dust linked to the bubble. The emission at 24 $\mu$m coincides with the inner part of the bubble, indicating warm dust inside. The spatial distribution of the emission in the far-IR from 70 to 160 $\mu$m coincides with the 8 $\mu$m bubble and the molecular emission, while the emission at 500 and 870 $\mu$m resembles that at 250 $\mu$m. The spatial distribution of the Herschel-PACS and Spire, and ATLASGAL emissions shows a cold dust component coincident with the molecular gas. Dust temperature determinations using the emissions at 250 and 350 $\mu$m allowed to estimate dust temperatures in the range 21-33 K for the cold dust component linked to the 8 $\mu$m bubble. 

Thermal radio continuum emission at 843 MHz and 1.4 GHz was detected from inside the bubble, indicating the existence of ionized gas and excitation sources, in agreement with the presence of warm dust. We conclude that a compact \hii\ region has developed. However, the identification of the exciting star is a difficult task and deserves additional studies.

A search for candidate YSOs was performed. We were able to identify many candidates in the VVV database projected onto the molecular clumps, although it is not clear whether all these candidates are linked to the molecular shell. This result suggests that star formation has been active recently. The \hii\ region is probably very young for the C\&C process to be active.

\acknowledgements 
C.E.C. acknowledges the kind hospitality of  M. Rubio and her family during her stay in Chile. 
V.F. acknowledges support from CONICYT Astronomy Program-2015 Research Fellow GEMINI-CONICYT (32RF0002) and from the Faculty of the European Space Astronomy Centre (ESAC), and would like to thank Ivan Valtchanov, Bruno Altieri, and Luca Conversi for their support and valuable assistance in Herschel data processing. 
We acknowledge the anonymous referee for very helpful comments.
The ATLASGAL project is a collaboration between the Max-Planck-Gesellschaft, the European Southern Observatory (ESO) and the Universidad de Chile.
This project was partially financed by CONICET of Argentina under project PIP 0356, UNLP under project 11/G120, and CONICyT of Chile through FONDECYT grant No. 1140839. We gratefully acknowledge use of data from the ESO Public Survey programme ID 179.B-2002 taken with the VISTA telescope and data products from the Cambridge Astronomical Survey Unit. Support for JB is provided by the Ministry of Economy, Development, and Tourism's Millennium Science Initiative through grant IC120009, awarded to The Millennium Institute of Astrophysics, MAS.
This work is based [in part] on observations made with the Spitzer Space Telescope, which was operated by the Jet Propulsion Laboratory, California Institute of Technology  under a contract with NASA. This publication makes use of data products from the Two Micron All Sky Survey, which is a joint project of the University of Massa chusetts and the Infrared Processing and Analysis Center/California Institute of  Technology, funded by the National Aeronautics and Space Administration and the  National Science Foundation.

\label{sec:refs}

\end{document}